\documentclass[pre,twocolumn,amsmath,amssymb,longbibliography,showpacs,superscriptaddress]{revtex4-1}
\usepackage{graphicx,color}
\usepackage{latexsym}
\usepackage{pgfplots}
\usepackage{empheq}

\newcommand{\caA}{{\mathcal A}}

\newcommand{\caC}{{\mathcal C}}
\newcommand{\C}{{\mathcal C}}
\newcommand{\caD}{{\mathcal D}}

\newcommand{\caR}{{\mathcal R}}

\newcommand{\caV}{{\mathcal V}}
\newcommand{\caW}{{\mathcal W}}

\newcommand{\dif}{{\rm d}}
\renewcommand{\vec}[1]{\boldsymbol{#1}}
\newcommand{\mean}[1]{{\left< #1 \right>}}

\begin{document}

\title{Inflow rate, a time-symmetric observable obeying fluctuation relations}

\author{Marco Baiesi}
\email{baiesi@pd.infn.it}
\affiliation{Department of Physics and Astronomy, University of Padova, Via Marzolo 8, I-35131 Padova, Italy}
\affiliation{INFN, Sezione di Padova, Via Marzolo 8, I-35131 Padova, Italy}

\author{Gianmaria Falasco}
\email{falasco@itp.uni-leipzig.de}
\affiliation{Institut f\"ur Theoretische Physik, Universit\"at Leipzig,  Postfach 100 920, D-04009 Leipzig, Germany}

\date{\today}

\begin{abstract}
While entropy changes are the usual subject of fluctuation theorems, we seek fluctuation relations involving time-symmetric quantities, namely observables that do not change sign if the trajectories are observed backward in time. We find detailed and integral fluctuation relations for the (time integrated) difference between  {\em entrance rate} and escape rate in mesoscopic jump systems. Such {\em inflow rate}, which is even under time reversal, represents the discrete-state equivalent of the phase space contraction rate. Indeed, it becomes minus the divergence of forces in the continuum limit to overdamped diffusion. This establishes a formal connection between reversible deterministic systems and irreversible stochastic ones, confirming that fluctuation theorems are largely independent of the details of the underling dynamics.
\end{abstract}


\maketitle

\section{Introduction}
The past two decades were characterized by the discussion of fluctuation relations (FRs), or fluctuation theorems, valid for systems arbitrarily far from equilibrium conditions~\cite{eva93,eva94,gal95,gal95b,jar97,jar97b,cro98,cro99,kur98,mae99,leb99,maes03,sei05,eva08,che06,har07,esp07,esp10,sag12,lee13,rah14} (the literature on FRs is vast, see more references in the reviews~\cite{eva02,ron07,mar08,sei12}). Initially the focus was on deterministic dynamical systems, where Lyapunov exponents were found to play a fundamental role~\cite{eva93,eva94,gal95,gal95b}. Although quantities as the phase space contraction rate (minus sum of all Lyapunov exponents) are related to the entropy production \cite{ron07}, this concept acquires a more immediate physical interpretation in terms of heat flows when stochastic systems are considered~\cite{jar97b,cro98,cro99,kur98,mae99,leb99,maes03,esp10,sei12,sag12,lee13,rah14}. For this reason, and because of the relevance for modern technological applications, nowadays it is more frequent to discuss FRs for stochastic dynamics.

The production of entropy, $S(\omega)$, being related to fluxes of heat, matter, etc., changes sign if one goes through the trajectory $\omega$ backward in time.  By now it is essentially understood that any FR under examination involves a form of entropy production, and that it yields a measure of the statistical asymmetry of physical processes in time. For example, the integral FR $\mean{e^{- S}} = 1$ ($\mean{\ldots}$ denotes a statistical average) and the convexity of the exponential function provide a statistical mechanical derivation of the second law of thermodynamics, $\mean{S}\ge 0$.
Such asymmetry in time exists already at the level of trajectories.
Entropy production is the physical quantity that always determines the time-antisymmetric sector of path probabilities~\cite{maes03,sei12}, 
\begin{equation}
\label{eq:P}
P(\omega)\sim e^{-\caA(\omega)} \sim e^{\frac 1 2 [S(\omega)-K(\omega)]},
\end{equation}
which is generally written as an exponential of some functional $\caA$. This includes also a time symmetric (TS) component $K(\omega)$. Yet, such a quantity is completely irrelevant in the standard procedure used to prove FRs, in which one compares path weights (\ref{eq:P}) with those of trajectories reversed in time. 

However, it is important to sharpen our understating of the meaning and the statistics of TS quantities, since it is becoming clear that they are necessary in the formulation of nonequilibrium statistical mechanics. A TS quantity that was recognized as a major player in characterizing the evolution of glassy systems is dynamical activity~\cite{lec05,mer05,mae06,lec07,gar07,gar09,gor09,mae14b}, which is just the number of jumps (or changes of state) that take place during a trajectory, regardless of their direction. Moreover, in the context of linear response theory, one finds fluctuation-dissipation relations where $K(\omega)$ complement the entropy production in determining the response of systems far from equilibrium~\cite{lip05,bai09,bai10,bai13,fal15b}. Mostly these TS observables were continued to be called dynamical activity, although sometimes also the names ``traffic''~\cite{mae08} or ``frenesy''~\cite{bai10} were adopted.
There are few examples of fluctuation symmetries for TS quantities~\cite{mae06,mae14b}. Therefore, to better understand the physics of nonequilibrium systems, there remains the interest of going deeper in this direction and find more variants of FRs for TS observables.

In this paper we briefly introduce some FRs for TS quantities. In discrete jump processes we study the {\em inflow rate} in a state, expressed as a properly defined entrance rate minus the standard escape rate. The derivation of FRs is based on an artificial ``auxiliary'' dynamics. A similar idea was recently put forward~\cite{rah14} to define the path-space probability of trajectories reversed in time when some transitions are only one way (i.e.~their reverse transition does not exists). With such an approach one may draw a generalized FR~\cite{rah14}. By taking this method to the extreme, we will define the auxiliary dynamics as that taking place when all jump rates are flipped. It is a simple mathematical choice that leads to specific FRs. The probability of the auxiliary dynamics, appearing in such FRs, can be traced out and one is left with integral FRs, valid for the normal dynamics and in terms of sound physical quantities. 

When we perform the limit to overdamped diffusive dynamics, we make contact with the approach by Maes and van Wieren~\cite{mae06}: the auxiliary dynamics here corresponds to flipping the force signs and the inflow rate becomes minus the divergence of the forces, which is the phase space contraction rate of the associated noiseless dynamics.  In our case the contraction rate is even under time reversal, differently from the case of reversible deterministic dynamical systems. Nevertheless, an integral fluctuation for this quantity is derived. Note that the contraction rate can be found, for conservative forces, within the definition of the so-called  effective potential, which is used in computations of reaction pathways~\cite{aut09} and in evaluations of the system activity~\cite{pit11,ful13}.

\section{\label{sec:jump}Jump process}

We consider a system with discrete states $\{\C\}$ whose probabilities $\rho_t$ evolve according to the master equation
\begin{equation}
\partial_t \rho_t(\C)= \sum_{\C'\ne \C} \left[ \rho_t(\C') k(\C' \to \C) -  \rho_t(\C) k(\C \to \C')\right],
\end{equation}
where $k(\C \to \C') $ is the transition rate from configuration $\C$ to $\C'$.
The last term in the master equation contains the probability $\rho_t(\C)$ times the escape rate from $\C$ 
\begin{equation}
\label{esc}
\lambda(\C) \equiv \sum_{\C'\neq \C}  k(\C \to \C').
\end{equation}
Here, next to this concept, we find it useful to define the entrance rate $\varepsilon(\C)$
as the sum of transition rates from $\C'$ to $\C$,
\begin{equation}
\label{entr}
\varepsilon(\C) \equiv \sum_{\C'\neq \C}   k(\C' \to \C).
\end{equation}
In physics often one considers transition rates obeying local detailed balance~\cite{kat83} at temperature $T=1/\beta$,
\begin{equation}
k(\C \to \C') \sim \exp\left[\frac \beta 2 \caW(\C\to\C')\right]
\end{equation}
where $\caW(\C\to\C')$ is the work done by the system on its environment 
(a heat bath in equilibrium). 
The term $\beta \caW(\C\to\C')$ thus represents the increase of entropy in the environment
associated to the jump $\C\to\C'$ in the system.

A trajectory from time $t_0=0$ to time $t$ is the time-ordered sequence 
$\omega\equiv \{\C(s)|0\le s \le t\} = \{\C_0=\C(0),\C_1,\C_2, \dots,  \C_n=\C(t)\}$ with $n$ jumps 
$\C_{i-1}\to \C_{i}$ taking place at times $t_i$.
For a system with initial density $\rho_0(\C)$, the probability to observe a trajectory $\omega$
is proportional to
\begin{align}\label{P}
P(\omega) \sim & \rho_0(\C_0) e^{- \lambda(\C_n) (t-t_n)}  
\prod_{i=0}^{n-1} k(\C_i\to \C_{i+1})  e^{- \lambda(\C_{i}) (t_{i+1}-t_i)} 
\nonumber\\
\sim & \rho_0(\C_0) e^{-\int_0^t d s \lambda(\C(s))}  \prod_{i=0}^{n-1} k(\C_i\to \C_{i+1})
\end{align}
(we are omitting a time-discretization prefactor that is common to all trajectories with the same number of 
jumps~\cite{mal15}).
The standard FR of the entropy production is obtained already at the level of single trajectories by comparing
$P(\omega)$ with the probability of the path reversed in time, which we obtain by applying an involution
$\theta$ that reverses the order of times ($\theta\omega\equiv \{\C(t-s)|0\le s \le t\}$, 
i.e.~the initial time of $\theta \omega$ is what it was the final time for $\omega$, etc.).
Thus,
\begin{eqnarray}\label{FRep}
\frac{P(\omega)}{P(\theta\omega)} = e^{S_{\rm tot}(\omega)},
\end{eqnarray}
with
\begin{eqnarray}
S_{\rm tot}(\omega) &=& - \ln \rho_t(\C_n) + \ln \rho_0(\C_0) + \beta \sum_{i=0}^{n-1}  
\caW(\C_i\to\C_{i+1}) \nonumber \\
 &=&  -\ln \rho_t(\C_n) + \ln \rho_0(\C_0)  + S(\omega),
\end{eqnarray}
where $S(\omega)$ is the entropy increase in the heat bath. 
From (\ref{FRep}) one readily observes that an increase of the entropy in the
bath is associated with processes more likely to take place in the normal direction of time, because typically
$\frac{P(\omega)}{P(\theta\omega)}>1$ in this case (excluding effects from the boundary terms).
Note that by construction this equation picks up the time-antisymmetric portion of the path measures 
and completely forgets about the integral of escape rates.

To obtain a new form of FR we do not simply consider time-reversal but we rather define an auxiliary dynamics
where all rates are replaced by the rates of the inverse transitions,
\begin{equation}
 k^*(\C \to \C') \equiv  k(\C' \to \C),
\end{equation}
so that the ``auxiliary'' escape rates correspond to the entrance rates of the normal dynamics,
\begin{equation}
\lambda^*(\C) \equiv \sum_{\C'\ne \C} k^*(\C \to \C') = \varepsilon(\C).
\end{equation}
Note that other choices for ``adjoint dinamics'' have already been discussed~\cite{che06,har07,esp07}. 
However, those are useful to single out specific entropy production terms only.

The state density $\rho^*$ with the auxiliary dynamics is chosen to be the same $\rho$ we
have with the normal dynamics.
Thus, the path probability under the auxiliary dynamics is
\begin{eqnarray}\label{P*}
P^*(\omega) &\sim& \rho^*_0(\C_0) e^{-\int_0^t \dif t' \lambda^*(t')}  \prod_{i=0}^{n-1} k^*(\C_i\to\C_{i+1})\nonumber\\
            &\sim& \rho_0(\C_0) e^{-\int_0^t \dif t' \varepsilon(t')} \prod_{i=0}^{n-1} k(\C_{i+1}\to\C_{i}),
\end{eqnarray}
while for the corresponding time-reversed path it reads
\begin{eqnarray}\label{P*theta}
P^*(\theta\omega) &\sim& \rho_t(\C_n) e^{-\int_0^t \dif t' \varepsilon(t')} \prod_{i=0}^{n-1} k(\C_{i}\to\C_{i+1}).
\end{eqnarray}
The initial densities of $P^*(\omega)$ and $P^*(\theta \omega)$ are taken to be the initial and final densities of the physical trajectory, respectively, $\rho_0$ and $\rho_t$. Any other choice is equally allowed, though (see Sec.~\ref{sec:contraction}).

It is clear that the ratio of the path measure~(\ref{P*}) or (\ref{P*theta}) with (\ref{P}) now yields the exponential
of a novel path-dependent term
\begin{eqnarray}
Y(\omega) &=& \int_0^t \dif t' [\varepsilon(\C(t'))-\lambda(\C(t'))] \nonumber\\
&\equiv&  \int_0^t \dif t' \, \caR(\C(t')),
\end{eqnarray}
in which there plays a crucial role the instantaneous inflow rate, 
i.e.~the imbalance between the entrance rate~\eqref{entr} and the escape rate~\eqref{esc}
\begin{equation}
\label{inflow}
\caR(\C) \equiv \varepsilon(\C) -  \lambda(\C).
\end{equation}
Specifically, the ratio of Eqs.~(\ref{P*}) and (\ref{P}) is
\begin{equation}
\label{Pstar}
\frac{P^*(\omega)}{P(\omega)}= e^{-S(\omega) - Y(\omega)},
\end{equation}
while the ratio  of Eqs.~(\ref{P*theta}) and (\ref{P}) is
\begin{equation}
\label{Pstar-theta}
\frac{P^*(\theta\omega)}{P(\omega)}= e^{\ln[ \rho_t(\C_n)/\rho_0(C_0)] - Y(\omega)}.
\end{equation}
Using the fact that $P(\theta \omega)$ can be normalized to $1$ ($\int \caD \omega P^*(\theta \omega) = 1=\mean{1}^*$)
and the conversion of statistical averages $\mean{\ldots}^* = \mean{\ldots P^*/P}$,
from (\ref{Pstar-theta}) we obtain the integral FR 
\begin{equation}\label{ft}
\mean{e^{\ln[ \rho_t(\C_n)/\rho_0(C_0)] - Y(\omega)}}=1,
\end{equation}
which becomes 
\begin{equation}\label{ftBis}
\mean{e^{-Y(\omega)}} \to 1 \qquad \textrm{for } t \to \infty
\end{equation}
if the boundary contribution is irrelevant. This occurs for instance 
for $t\to \infty$ if the system has a finite number of states
and is in a stationary regime, $\rho_t=\rho_0 \, \forall t$.
The convexity of the exponential yields also the inequality $\mean{Y(\omega)} \ge 0 $. A positive inflow rate 
is to be expected as it is easier on average to jump into the states with high $\rho$ than into those with low 
$\rho$.

With a similar procedure but using 
(\ref{Pstar}) rather than (\ref{Pstar-theta}) one obtains another FR 
\begin{equation}\label{ft2}
\mean{e^{-Y(\omega)-S(\omega)}}=1,
\end{equation}
where the entropy production $S$ reappears next to $Y$.
Note that~\eqref{ft} and~\eqref{ft2} are valid also in transient conditions. 
An equation as (\ref{ft2}) might be particularly interesting in transient regimes because it does
not contain explicitly the initial and final density of states. Possibly it might help to
study glassy systems, which are the paradigm of transient dynamics.

\section{\label{sec:diff}Overdamped diffusion}

In order to identify the transformation of a diffusive dynamics equivalent to inverting the rates 
in a jump process we start from the generic Fokker-Planck equation~\cite{ris89} for the density $\rho_t(x)$,
\begin{equation}\label{fpe}
\partial_t \rho_t(x)= - \partial_x (\mu F(x)\rho_t(x)) + D \partial^2_x \rho_t(x),
\end{equation}
where $\mu$ is a mobility, $D$ is a diffusion constant, and $F(x)$ is a force.
We discretize phase space in tiny units of 
size $\delta$, so (\ref{fpe}) turns into a master equation
for configurations $\C=\ldots,x-\delta,x,x+\delta,\ldots$. With the standard
assumption that the dynamics is performed by random walks with jumps to nearest neighbors $x\to y=x\pm \delta$,
the discretization yields 
\begin{eqnarray}\label{foe}
\partial_t \rho_t(x)&=& 
- \frac{\mu}{2\delta} [F(x+\delta)\rho_t(x+\delta)- F(x-\delta)\rho_t(x-\delta)] 
\nonumber\\
&&+ \frac{D}{\delta^2} [\rho_t(x+\delta) + \rho_t(x-\delta)-2\rho_t(x)],
\end{eqnarray}
that can be rewritten as a master equation with transition rates
\begin{equation}
k(x\to y)= \left\{ \begin{array}{ll}
\displaystyle \frac{D}{\delta^2} - \frac \mu {2\delta} F(x-\delta)  & \textrm{if $y=x-\delta$}\\
  & \\
\displaystyle \frac{D}{\delta^2} + \frac \mu {2\delta} F(x+\delta)  & \textrm{if $y=x+\delta$}.
  \end{array} \right.
\end{equation}
When we apply the transformation $k^*(x\to y)= k(y\to x)$ we obtain
\begin{equation}
k^*(x\to y)=  \left\{ \begin{array}{ll}
\displaystyle \frac{D}{\delta^2} + \frac \mu{2\delta}  F(x)  & \textrm{if $y=x-\delta$}\\
  & \\
\displaystyle \frac{D}{\delta^2} - \frac \mu {2\delta} F(x)  & \textrm{if $y=x+\delta$}.
  \end{array} \right.
\end{equation}
The difference in the state where the forces are evaluated is of order $O(1)$~\cite{van11}, and so vanishes when we go back to the continuous limit $\delta\to 0$  ($D$ is assumed to be a constant). We conclude that changing the sign of $F$ (or alternatively of $\mu$) gives the transformed rates in terms of the original ones, namely $k^*_F(x\to y)=k_{-F}(x\to y)$.
The path weight associated to (\ref{fpe}), describing overdamped diffusive dynamics, is 
\begin{align}
P(\omega) \sim \rho_0(x(0)) \exp\bigg\{ 
& -\int_0^t \dif t' \frac{[\dot x(t') - \mu F(t')]^2}{4D} \nonumber\\
& - \frac \mu 2 \int_0^t \dif t' \partial_x F(t')\bigg\}
\end{align}
and it is straightforward to obtain the equivalent of (\ref{ft}) by applying time reversal together with the transformation $F^*= -F$ (or equivalently $\mu^* = -\mu$), which again gives the probability ratio 
\begin{equation}
\label{Pratio-rw}
\frac{P^*(\theta\omega)}{P(\omega)} = \frac{\rho^*_t(x(t))}{\rho_0(x(0))} e^{-Y(\omega)}
 =e^{\ln[ \rho_t(x(t))/\rho_0(x(0))] - Y(\omega)},
\end{equation}
where now the integral $Y$ is
\begin{equation}
\label{Y-rw}
Y(\omega) = - \mu \int_0^t \dif t' \partial_x F(t').
\end{equation}
In~\cite{mae06} we found the only previous example where one force was flipped to get a FR for TS quantities.
Alternatively, we can get again
\begin{equation}
\label{Pstar2}
\frac{P^*(\omega)}{P(\omega)}= e^{-S(\omega) - Y(\omega)},
\end{equation}
where the entropy increase in the heat bath is here defined by
\begin{equation}
S(\omega)= \frac{\mu}{D} \int_0^t \dif t' F(t') \dot x(t').
\end{equation}
Therefore, we find Eqs.~(\ref{ft}),~(\ref{ftBis})~and~(\ref{ft2}) to be valid also in overdamped diffusing systems. Besides, from \eqref{Pratio-rw} one can as well derive detailed FRs that link the statistics of observables (odd under the joint inversion of time and forces) in two systems subject to opposite forces. Examples of practical interest are systems of non-interacting particles exposed to controllable external fields. 

In more than one dimension and with $\mu=1$, Eq.~(\ref{Y-rw}) reads
\begin{equation}
\label{Y-rw-d}
Y(\omega) =  -\int_0^t \dif t'  \partial_{x_i}  {F_i} (\vec x(t'))\,.
\end{equation}
Einstein notation is used from here onward and the vector state is denoted as $\vec x \equiv \{x_i\}$. 
For overdamped stochastic systems the inflow rate is thus represented by a function with the structure of a divergence
of forces,
\begin{equation}
\label{R-rw-d}
\caR(\vec x) =  - \partial_{x_i}  {F_i} (\vec x)
\end{equation}
when all mobilities are equal to $1$.
This form has an analogous version in deterministic evolution, as discussed in the next section.

\section{\label{sec:contraction}Analogy with the contraction rate}

Consider a dynamical systems composed of $i=1,\ldots,N$ degrees of freedom evolving according to deterministic equations
\begin{equation}
\label{ds}
\dot x_i = F_i(\vec x).
\end{equation}
In phase space the system is described by a density $\rho$ which evolves according to the continuity equation, enforcing probability conservation:
\begin{align}
\partial_t \rho(\vec x(t),t) &= - \partial_{x{_i}}  [\dot {x}_i(t) \rho(\vec x(t), t)]\nonumber\\
&= - \partial_{x{_i}}  [F_i(\vec x(t)) \rho(\vec x(t), t)].
\end{align}
Rearranging the terms and introducing the Lagrangian derivative $\dif/ \dif t \equiv \partial_t + \dot x_i \partial_{x_{i}}$ accounting for time variations along trajectories, we find
\begin{equation}\label{Liouville}
\frac{\dif }{\dif t} \ln \rho(\vec x(t),t)=  - \partial_{x{_i}} F_i(\vec x(t)) \equiv \sigma(\vec x(t)).
\end{equation}
In (\ref{Liouville}) we introduced the instantaneous contraction rate $\sigma(\vec x)$, which measures the logarithmic rate of contraction of phase space volumes \cite{chaos08} ---it is identically zero in Hamiltonian systems and on average positive in dissipative ones \cite{gas15}. This is exactly the quantity~(\ref{R-rw-d}) for which we have derived a FR for the diffusive dynamics. Moreover, since the transformation used to find the FR in the discrete dynamics (see Sec.~\ref{sec:jump}) is analogous to that used to derive the FR in the diffusive dynamics, we conclude that the inflow rate $\caR(\C)$ is the equivalent
of the contraction rate $\sigma(\vec x)$.

There is indeed a procedure that illustrates this correspondence. In a discrete state system, we spread uniformly the occupation probability on the
configuration $\C$ and on its neighbors, i.e. those $\C'$ such that $k(\C \to \C') \neq 0$. At $t=0$ we thus have $\rho_0(\C) = \rho_0(\C') = 1/\caV(\C)$, where $\caV(\C)$ is the number, or ``volume'', of the states centered around $\C$. Shortly after the preparation, the time derivative of the Boltzmann entropy associated to this locally flat density is expected to give the logarithmic variation of the inverse volume centered on $\caC$. Therefore we have
\begin{align}
&\partial_t \ln \rho_t(\C(t)) |_{t=0}  =  \caV(\C) \partial_t \rho_t(\C) |_{t=0}\nonumber\\
&\quad =  \caV(\C)  \sum_{\C'} [\rho_t(C') k(\C'\to \C) - \rho_t(C) k(\C\to \C') ] \bigg|_{t=0}\nonumber\\
&\quad =  \sum_{\C'} [ k(\C'\to \C) - k(\C\to \C') ]
\nonumber\\
&\quad =  \caR(\C),
\end{align}
which shows that the inflow rate measures the ``volume" contraction rate. 

The phase-space contraction rate is a key ingredient in the derivation of FRs for deterministic dynamics~\cite{eva02,ron07}. Hence we deem it interesting to strengthen the correspondence between thermostated deterministic systems (i.e., Hamiltonian systems with an added nonlinear friction term) and stochastic ones applying the ideas of Sec.~\ref{sec:jump} and Sec.~\ref{sec:diff} to underdamped diffusion, where the phase space does not simply reduce to the configuration space. For simplicity we consider two conjugated degrees of freedom only, $\vec x=\{q,mv\}$, exemplifying a particle of mass $m$ and friction coefficient $\gamma$ moving in a force field $F$. The motion is described by the Langevin equations
\begin{align}
\dot q= v, \quad m\dot v= - \gamma v + F(q) +\sqrt{2 D_v} \xi,
\end{align}
where $\xi$ is a standard Gaussian white noise with unit variance. Since the associated path weight is 
\begin{align}\label{P_under}
P(\omega) \sim &\rho_0(x(0),v(0)) e^{ \frac \gamma 2 t } \nonumber\\
&\exp\bigg\{  -\int_0^t \dif t' \frac{[m\dot v(t') + \gamma v(t') -F(t')]^2}{4D_v}\bigg\},
\end{align}
the time-integrated contraction rate of the noiseless dynamics, $Y(\omega)= \gamma t$, is singled out by comparison with an  auxiliary dynamics having negative friction coefficient, i.e. $\gamma^*=-\gamma$. Recalling that time reversal here implies $\theta v(t')=-v(t-t')$, the analogous of \eqref{Pratio-rw} for underdamped diffusion is
\begin{equation}\label{Pstartheta_u}
\frac{P^*(\theta\omega)}{P(\omega)} 
 =e^{\ln[ \rho^*_t(x(t),-v(t))/\rho_0(x(0),v(0))] -\gamma t}.
\end{equation}
Essentially, this auxiliary dynamics emulates the time-reversal transformation of thermostated Hamiltonian systems. There~\cite{eva90}, the friction coefficient $\gamma$ is indeed replaced by  the thermostat multiplier, that is an odd function of the velocity $v$ and thus changes sign upon time inversion.  Moreover, this auxiliary dynamics is equivalent to that used in Sec.~\ref{sec:diff} for underdamped diffusion, since $\gamma=1/\mu$ and we have already noticed that flipping the forces is the same as changing the sign of the mobility.

Similarly to the deterministic case, equations~(\ref{Pratio-rw})~and~(\ref{Pstartheta_u}) can be turned into an integral FR for the dissipation function $\Omega$ \cite{eva02},
\begin{equation*}
\int_0^t \dif t' \Omega(\vec x(t')) \equiv \ln \rho_0(\vec x(0)) - \ln \rho_0(\vec x(t)) + \int_0^t \dif t' \sigma(\vec x(t')),
\end{equation*}
exploiting the freedom in choosing the initial density of the backward auxiliary trajectory. Specifically, taking $\rho_t^*=\rho_0$ in (\ref{Pratio-rw}) we obtain the nonequilibrium partition identity, i.e. the integral version of the FR~\cite{eva08}  
\begin{equation}
\mean{\exp\left( -\int_0^t \dif t' \Omega(t') \right)}=1,
\end{equation}
valid in overdamped diffusing systems for arbitrary initial densities and nonconservative forces.
The same relation is obtained for underdamped diffusion from \eqref{Pstartheta_u}, provided that $\rho_0$ is independent of the sign of $v$, as it is, e.g., at equilibrium.

Finally, it is worth noting that the quantity (\ref{Y-rw-d}) can also be related to the (finite-time) Lyapunov exponents $\Lambda_i$ of the system. The latter are defined by considering the growth rates of $k$-dimensional volumes supported by $k$ linearly independent perturbations $\delta \vec x^{(i)}$, with $1 \leqslant k \leqslant N$ \cite{rue85}:
\begin{equation}\label{Lyapunov}
\textrm{Vol}_k(\{\delta \vec x^{(i)}(t)\}) \sim e^{\sum_{i=1}^k  \Lambda_i t}\,.
\end{equation}
The time evolution of such a perturbation $\delta \vec x$ in the initial condition of (\ref{ds}) is 
\begin{equation}
\label{dx}
\delta \dot{x}_i = {F_i}(\vec x + \delta{ \vec x} ) - {F_i}(\vec x )
\approx (\partial_{x_i} {F_j}) \delta{x_j} 
\equiv (\nabla F)_{ij} \delta{x_j} ,
\end{equation}
which has the solution
\begin{equation}
\label{xt}
\delta {x}_i(t) =
\exp\left(\int_0^t \nabla  F(t') \dif t' \right)_{ij} 
 \delta {x}_j(0).
\end{equation}
Therefore, using the formula relating the determinant to the trace,
$\det( \exp A) = \exp ({\rm Tr} A)$, the relative variation of the $N$-volume is obtained as
\begin{align}
 \det\left( \frac{\delta {x_i}(t)}{\delta {x_j}(0)}\right)
&= \exp\left(\int_0^t  \partial_{x_i}  F_i(t') \dif t' \right)\nonumber\\
&= \exp\left(-\int_0^t  \sigma(t') \dif t' \right)\,,
\end{align}
which gives the time-averaged contraction rate in terms of the negative sum of all Lyapunov exponents,
\begin{equation}
\frac 1 t \int_0^t \sigma(t') \dif t' = -\sum_{i=1}^N \Lambda_i.   
\end{equation}
If (\ref{ds}) is stochastic, as in the overdamped diffusion considered in the previous section, the Lyapunov exponents identify the separation rate of nearby trajectories subjected to the same noise realization, and Eqs.~(\ref{Pratio-rw})~and~(\ref{Pstar2}) are statements about the statistics of their spectrum. Note that Eqs.~(\ref{Pstar2})~and~(\ref{Pratio-rw}) can be restricted to quantities measured on subspaces of the full phase space using auxiliary dynamics where only some components of the forces are changed in sign. In particular, when $F_i^*=-F_i$ if $i=k$ and $F_i^*=F_i$ otherwise, we get for example (no Einstein summation)
\begin{equation}\label{Yk}
\mean{\exp\left[-\frac 1 D \int_0^t \dif t'  F_k(t') \dot x_k(t')+ \int_0^t \dif t' \partial_{x_k}F_k(t') \right]}=1.
\end{equation}
Since a $k$-volume is obtained as the determinant of the $k \times k$ minor of $\delta x_i(t)/\delta x_j(0)$, Eq.~(\ref{Yk}) relates a single Lyapunov exponent $\Lambda_k$,
\begin{equation}
e^{ t \Lambda_k}= \textrm{Vol}_k(t)-\textrm{Vol}_{k-1}(t)= \exp \left( \int_0^t \dif t' \partial_{x_k}F_k(t') \right),
\end{equation} 
to the entropy produced in the $k$ direction.

\section{Conclusions}
We have introduced the concept of {\em inflow rate}. The fluctuations of its time integral are characterized by two fluctuation relations, one involving also boundary terms (density of states) and one involving the entropy production. In the continuum limit bringing jump systems to overdamped diffusion, the inflow rate becomes (minus) the divergence of forces, i.e.~the phase space contraction rate. This quantity in reversible deterministic dynamical systems is known to be odd under time reversal. However, in general the phase space contraction rate can have any or no parity. In particular, in our case it results even under time reversal, due to the fact that the overdamped Langevin dynamics is not time reversible. Note that time reversibility is a condition commonly invoked in the derivations of FRs, hence it is sufficient, but it is known to be unnecessary~\cite{col11b,col12b,col13}.

We hope that these findings will be useful in developing nonequilibrium statistical mechanics ---e.g. in deriving symmetry properties of response coefficients far from equilibrium, as much as standard FRs are employed around equilibrium to obtain the Onsager reciprocity relations \cite{gal96}--- for which by now it is clear that we need not only to have entropy flows under control, but also to better understand the statistics of activities that have an even parity under time reversal.

\section*{Acknowledgments} We thank R.~L.~Jack, A.~Lazarescu, S.~Lepri, and C.~Maes for useful discussions,
and L. Rondoni for important clarifications.


%

\end{document}